\documentclass[12pt]{article}
\usepackage{amssymb}
\usepackage[dvips]{graphics}
\usepackage[dvips]{color}
\usepackage{graphicx}
\begin{document}
\begin{center}
{\large\bf Evading the astrophysical limits on light pseudoscalars}
	
\bigskip
Pankaj Jain and Subhayan Mandal\\

Physics Department\\
I.I.T. Kanpur\\
Kanpur, India 208016\\
email: pkjain@iitk.ac.in, shabsslg@iitk.ac.in

\end{center}

\bigskip
\noindent
{\bf Abstract:} We study the possibility of evading astrophysical 
bounds on light pseudoscalars. 
We argue that the solar bounds can be evaded if we have a sufficiently
strong self coupling of the pseudoscalars. 
The required couplings do not 
conflict with any known experimental bounds. 
We show that it is possible to 
find a coupling range such that the results of the recent PVLAS experiment
are not in conflict with any astrophysical bounds.

\bigskip

\section{Introduction}
In a recent paper, Zavattini  et al. \cite{Zavattini} have reported a 
rotation of polarization
of light in vacuum in the presence of a transverse magnetic field. If
we interpret this rotation in terms of the coupling of a  
light pseudoscalar particle to photons,
\begin{equation}
{\cal L}_{\phi\gamma\gamma} = {1\over 4 M_\phi} \phi F_{\mu\nu}\tilde 
F^{\mu\nu}\ ,
\label{Lphigg}
\end{equation}
we find that the allowed range of parameters is 
	$1\times 10^5\ {\rm GeV} \le M_\phi \le 6\times 10^5\ {\rm GeV}$ and
	$0.7\ {\rm meV} \le m_\phi \le 2\ {\rm meV}$ \cite{Zavattini}, 
where $m_\phi$ is the mass
of the pseudoscalar. This range of parameters does not conflict with 
any laboratory bounds \cite{Cameron}. However these values are ruled out by 
astrophysical considerations \cite{astrobounds} if we assume that the 
pseudoscalar is an axion. 
For a review of axion physics see, for example, Ref. \cite{review}. 
The most stringent astrophysical
limit comes from SN1987A which suggests
that the mass of the axion cannot be greater than 0.01 eV \cite {PDG}. 
Assuming that this particle is an invisible axion, this in turn
implies an upper bound on the coupling, 
$g_{\phi\gamma\gamma} < 2 \times 10^{-12} \rm{GeV}^{-1}$,
upto a model dependent factor of order unity.
	The limit on the axion mass arises by demanding that axion emission
should not lead to too much energy loss from the core. Similar but less
stringent limits can be obtained by considering energy loss from the
core of the sun. The coupling, Eq. \ref{Lphigg},
also leads to several interesting astrophysical polarization effects
\cite{polarization}.
For standard axion, its mass and coupling to photons
are both related to the Peccei-Quinn scale \cite{PQ}. 
It is clearly of interest to see if the astrophysical bounds can somehow be 
evaded \cite{Ringwald}. All the astrophysical bounds assume that the pseudoscalar 
couplings are so small that once produced it will freely escape from the
source, which may be the sun or a red giant or a supernova. 
In the present paper we examine whether the pseudoscalar particles
might be trapped inside the sun. 

\section{Trapping Pseudoscalars}

	We assume the following interaction lagrangian of pseudoscalars, 
	\begin{equation}
	{\cal L}_I = {1\over 4 M_\phi} \phi F_{\mu\nu}\tilde
	        F^{\mu\nu} + {\lambda\over 4! } \phi^4
	\label{eq:Lphi_I}
	\end{equation}
Here we have also included the self coupling of the pseudoscalars. 
The value of the pseudoscalar coupling to photons will be 
taken from data \cite{Zavattini}. 
We do not assume $\phi$ to be an axion and use only experimental data
to determine its couplings. We point out that
the self couplings are  completely 
unconstrained experimentally.

\begin{figure}
	\begin{center}
	\includegraphics[scale=1.0]{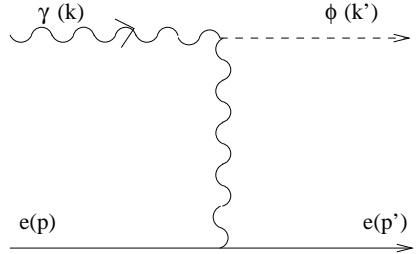}
	\end{center}
	\caption{Feynman diagram for the 
	Primakoff  process for the conversion of photons $\gamma(k)$
	into pseudoscalars $\phi(k')$.  
	 }
	\label{fig:primakoff}
	\end{figure}

\begin{figure}
	\begin{center}
	\includegraphics[scale=1.0]{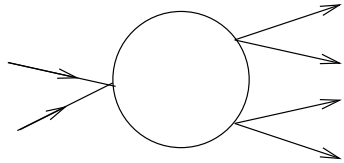}
	\end{center}
	\caption{The process $\phi\phi\rightarrow \phi\phi\phi\phi$ through
	a loop diagram. }
	\label{fig:loop}
	\end{figure}

We propose the following mechanism for trapping pseudoscalars. We assume
that the self coupling $\lambda$ is of order unity. 
The photons in the sun's core
convert into pseudoscalars due to the Primakoff process 
shown in 
fig. \ref{fig:primakoff}. The temperatures are such that almost the
entire flux of photons emerging from the core is converted into 
pseudoscalars. The pseudoscalars begin accumulating inside the core due 
to scattering on other pseudoscalars through the process $\phi\phi
\rightarrow \phi\phi$. This is possible if the self 
coupling of pseudoscalars $\lambda$ is sufficiently strong. 
We can also have a significant cross section for the process 
$\phi\phi\rightarrow \phi\phi\phi\phi$ through the loop diagram, fig.  
\ref{fig:loop}, if the 
coupling $\lambda$ is strong. This process 
leads to a degradation of the energy per particle for pseudoscalars. Hence
as the pseudoscalars accumulate they start producing pseudoscalars of
lower energy through this process and their density increases rapidly.
This process stops when the energy per particle approaches the mass
of the pseudoscalars. As we show the final density of pseudoscalars
is sufficient to trap pseudoscalars inside the star. We also find that 
the mean free path of pseudoscalars becomes so small that
radiative transport \cite{Raffelt} occurs primarily through photons.  
	
	\begin{figure}
	\begin{center}
	\includegraphics[scale=1.0]{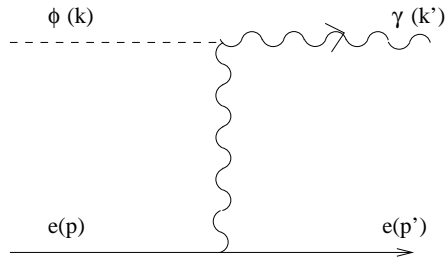}
	\end{center}
	\caption{Feynman diagram for the inverse 
	Primakoff process for the conversion of pseudoscalars
	$\phi(k)$ to photons $\gamma(k')$. }
	\label{fig:primakoff1}
	\end{figure}

We first compute the pseudoscalar density in the core of the sun assuming
the current equilibrium conditions. We assume the core radius to be one-fifth
of the solar radius $R_\odot$. The mean density in this region is approximately
60 gm/cm$^3$. The average temperature in the sun's core is approximately 
$1.7 \times 10^{7}\ {\rm K}$. The 
pseudoscalars are produced by the Primakoff like process shown in fig. 
\ref{fig:primakoff} and have mean energy roughly equal to 800 eV. We assume 
that the production rate of photons inside the
core is equal to their rate of conversion into pseudoscalars. 
This is a reasonable assumption
since the energy loss due to pseudoscalar production by this process is 
found to be larger than luminosity of sun. The solar luminosity is $3.9\times
10^{33}$ erg/s. Hence the mean photon energy production rate
inside the sun is roughly 2 erg/(gm sec). It is clear that the production rate
will be larger inside the core due to its higher temperature. 
For a conservative estimate we take the 
photon energy production rate inside the core to be 1 erg/(gm sec). The 
mean thermal energy per photon inside the core is roughly 800 eV ($\approx
1.3\times 10^{-9}$ ergs). Hence the number of photons produced per second is
approximately $5.21\times 10^{41}$. Since we have assumed that
all the photons produced convert into pseudoscalars we find that this is
also the number of pseudoscalars produced per second. Assuming that these 
particles escape freely from the sun we find that the flux of pseudoscalars
from the sun's core is $2.14\times 10^{20} {\rm cm}^{-2}{\rm s}^{-1}$. 
If the pseudoscalars are in a steady state the number
density of pseudoscalars inside the core is $7.1\times 10^9 {\rm cm}^{-3}$.

We next determine the mean free path of a pseudoscalar 
propagating through this medium. The cross section of the process 
$\phi(p_1)\phi(p_2)\rightarrow \phi(p_3)\phi(p_4)$ 
due to the $\phi^4$ coupling at leading order in perturbation theory
is given by
\begin{equation}
\sigma_{\phi\phi} = {\lambda^2\over 32\pi E_{\rm cm}^2} 
 =  9.94 \times 10^9 \lambda^2\left[{10^{-6}\ {\rm 
GeV}\over E_{cm}}\right]^2\ {\rm GeV}^{-2} 
\end{equation}
With $\lambda=1$ and $E_{cm} = 1$ KeV, the mean free path $l_{\phi\phi}
=1/(n_\phi \sigma_{\phi\phi})=3.6\times 10^7$ cm. This is almost three
orders of magnitude smaller than the core radius $0.2R_\odot= 1.4\times 10^{10}
$ cm. This implies that due to self interaction pseudoscalars will not
be able to escape freely from the sun but will instead start accumulating
inside the core. Hence our initial assumption that the pseudoscalars escape 
freely from the sun is not valid.  

We next consider the contribution due to higher order processes. We focus
on the process $\phi\phi\rightarrow\phi\phi\phi\phi$ for which
the leading order diagram is shown in fig. \ref{fig:loop}.
This process is suppressed compared to the leading order process by
two powers of $\alpha_\lambda = \lambda^2/4\pi$. With $\lambda=1$, this
process is down only by a factor of 100 and can contribute significantly.
This process leads
to fragmentation of $\phi$ particles and hence will degrade their energy
per particle. The fragmentation stops when the energy per particle approaches
their mass, which is of order 1 meV. Since the initial energy is 1 KeV, this
will lead to an increase in the number of particles by a factor of $10^6$
and hence will result in a considerable enhancement of the number density
of $\phi$ particles. 
The center of mass energy is now
equal to $\sqrt{2m_\phi E_\phi}$. This further enhances the cross section for
the process $\phi\phi\rightarrow\phi\phi$ which becomes $2.5
\times 10^{15}$ GeV$^{-2}$. 

The above analysis shows that, during the main sequence phase, the pseudoscalars
will accumulate inside the sun due to self interaction. As they accumulate,
processes which deplete pseudoscalars also start contributing significantly.
The dominant process contributing to depletion is the inverse Primakoff
process shown in fig. \ref{fig:primakoff1}. The mean energy of the pseudoscalars
is much smaller than that of photons. Hence the photon produced by the inverse
Primakoff process will have much lower energy, of the order of meV. The photons
produced will be quickly absorbed by electrons by the inverse Bremsstrahlung
process and thermalize in the medium.
Eventually the energy deposited into pseudoscalars will equal the energy 
released by pseudoscalars due to their conversion back into photons. 
We next determine the density profile of the pseudoscalars in 
this equilibrium situation. 

The cross section for the inverse process (fig. \ref{fig:primakoff1}) 
is approximately $10^{-12}$ GeV$^{-1}$ for $E_\phi<< m_e$, 
assuming the target particles to be electrons.
Their number density will reach 
steady state only when the rate at which energy is gained by the pseudoscalars 
becomes equal to the rate at which they loose energy. The mean energy of 
photons is about $10^6$ times larger than that of pseudoscalars. Hence 
a pseudoscalar produced by 
the primakoff like process (fig. \ref{fig:primakoff}) will have roughly $10^6$
times the energy of the photon produced by the inverse
process (fig. \ref{fig:primakoff1}). The production rate of pseudoscalars
per unit volume is $\sigma_{\gamma X\rightarrow \phi X}n_\gamma n_X v$ 
where $v=c$ is the relative velocity and $n_\gamma$ is the photon density
of the medium. We set this equal to $10^{-6}$ times
the conversion rate of pseudoscalars
into photons $\sigma_{\phi X\rightarrow \gamma X}n_X n_\phi v$. This
gives us the final number density of pseudoscalars when the system reaches
steady state $n_\phi = 10^6 n_\gamma \sigma_{\gamma X\rightarrow \phi X}/
\sigma_{\phi X\rightarrow \gamma X}$. The number density of photons in 
the core is roughly $10^{23}$ per cm$^3$. Since the two cross sections
are approximately equal, we find that the pseudoscalar number
density required to achieve steady state is roughly $10^{29}$ per cm$^3$. 
This leads to a mean
free path of pseudoscalars of order $10^{-17}$ cm inside the core of the
sun. This is much smaller than the mean free path of photons inside
the core. Hence the contribution of $\phi$ particles to radiative transport
inside sun would be negligible compared to the contribution of photons,
once the sun becomes a main sequence star.

\begin{figure}
\begin{center}
\includegraphics[scale=1.0,angle=0]{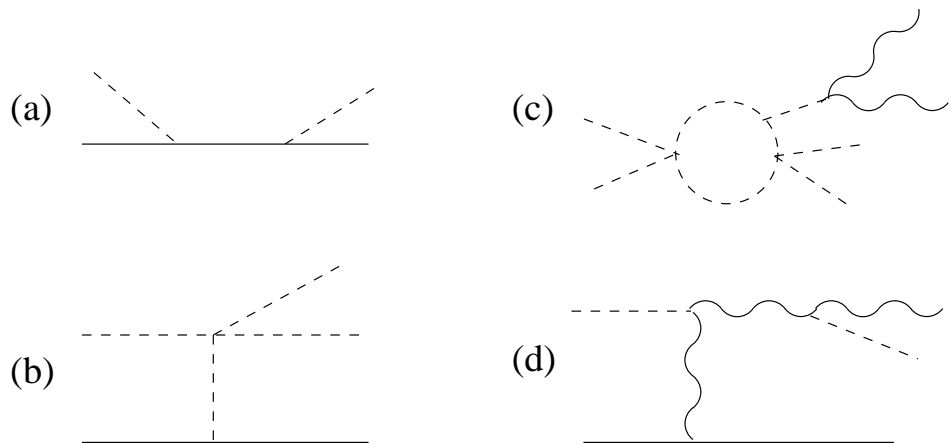}
\end{center}
\caption{Some of the diagrams which contribute to energy loss of pseudoscalars
as they propagate through sun. Here the pseudoscalar, electron and photon are
represented by dashed, solid and wavy lines respectively.}
\label{fig:energyl}
\end{figure}

So far we have only considered pseudoscalars inside the sun's core assuming
steady state conditions. We expect the pseudoscalar density to be significant
outside the core, extending much beyond the solar radius.   
This pseudoscalar halo will be gravitationally 
bound to the sun. We assume it to be spherically symmetric since we expect
its rotational speed to be small. Due to the fragmentation process 
the pseudoscalar kinetic energies after emerging from the core 
can be atmost as large as their mass. Hence their velocity may still be 
comparable to the velocity of light, which is much larger in comparison
to the escape velocity from the surface of the sun, $v_\odot\approx
6\times 10^5$ m/s. 
We assume that the
pseudoscalars loose energy as they propagate inside the sun such that their
velocities 
become nonrelativistic. There are many processes by which they
can loose energy. For this we assume additional terms in
the interaction lagrangian
\begin{equation}
\Delta{\cal L}_I = g_{_A}\phi\bar \psi \gamma_5 \psi + \lambda_1\phi^3 \ . 
\label{eq:DelLphi_I}
\end{equation}
where the term proportional to $\lambda_1$ is parity violating. We are allowed
to add such terms since parity is not a symmetry of nature. The coupling 
$g_A$ is also constrained by astrophysical processes. 
However, due to pseudoscalar self coupling and 
accumulation inside
the sun the limits on $g_{_A}$ are also much less stringent in comparison to 
the limit, $g_{_A} <0.80 \times 10^{-10}$ \cite{Frieman}, imposed by assuming
that the pseudoscalar is an invisible axion. 
Laboratory bounds
on this coupling also exist if we
assume a very small mass of the pseudoscalar, i.e. 
if $ m_{\phi} \le 10^{-6} ~{\rm eV} $  \cite {Vorobyov}.
The coupling $\lambda_1$
is unconstrained. Some of the processes which can contribute to energy loss are
shown in fig. \ref{fig:energyl}. The diagram, fig. \ref{fig:energyl}a, 
by itself can contribute significantly
to energy loss if the coupling $g_A\sim 10^{-7}$, which is not ruled out by
any laboratory experiments. Furthermore the diagram, fig. \ref{fig:energyl}b, 
is only first order in the coupling $g_A$ and may lead to  
sizeable energy loss due to scattering on non-relativistic electrons. 

We may estimate the density profile of the pseudoscalar halo  for $r>R_\odot$
by imposing steady state
conditions. The pressure of pseudoscalars at distance $r$ from center of
sun is $P=\rho_\phi <v^2>/3$. The mean velocity at distance $r$ is given by
\begin{equation}
<v^2> = c + {2 G(M_\odot + M_r)\over r}
\label{eq:v2}
\end{equation}
where $c= <v_0^2> - 2G(M_\odot + M_{R_\odot})/R_\odot $ is a constant, 
$M_\odot$ is the mass of the sun and
$M_r$ is the total mass of pseudoscalar particles within a sphere of
radius $r$ centered at the center of sun. We also have
\begin{equation}
{d M_r\over dr} = 4 \pi r^2\rho_\phi(r)\ , 
\label{eq:dMr}
\end{equation}
and
\begin{equation}
{d P\over dr} = -  {G (M_\odot + M_r)\rho\over r^2} . 
\label{eq:dPr}
\end{equation}
Using equations \ref{eq:v2}, \ref{eq:dMr} and \ref{eq:dPr} and the equation
$P=\rho_\phi <v^2>/3$, we find, 
\begin{equation}
{d \rho_\phi\over dr} = -{1\over <v^2>}  \left[{G (M_\odot + M_r)\rho_\phi
\over r^2} + 8\pi G\rho_\phi^2 r\right]. 
\label{eq:dRhor}
\end{equation}
Since the right hand side is negative for all $r$, we find that $\rho_\phi$
decreases with $r$. 
In the small $r$ regime where $M_r<< M_\odot$ and $8\pi \rho_\phi r^3<<
M_\odot$ we find that 
$\rho_\phi(r)/ \rho_\phi(r_0) \approx \sqrt{v^2(r)/ v^2(r_0)}$.  
In the large $r$ regime, $M_r>> M_\odot$ we find that $\rho_\phi
\propto 1/r^2$ and hence $M_r\propto r$. 

\begin{figure}
\begin{center}
\includegraphics[scale=1.0,angle=0]{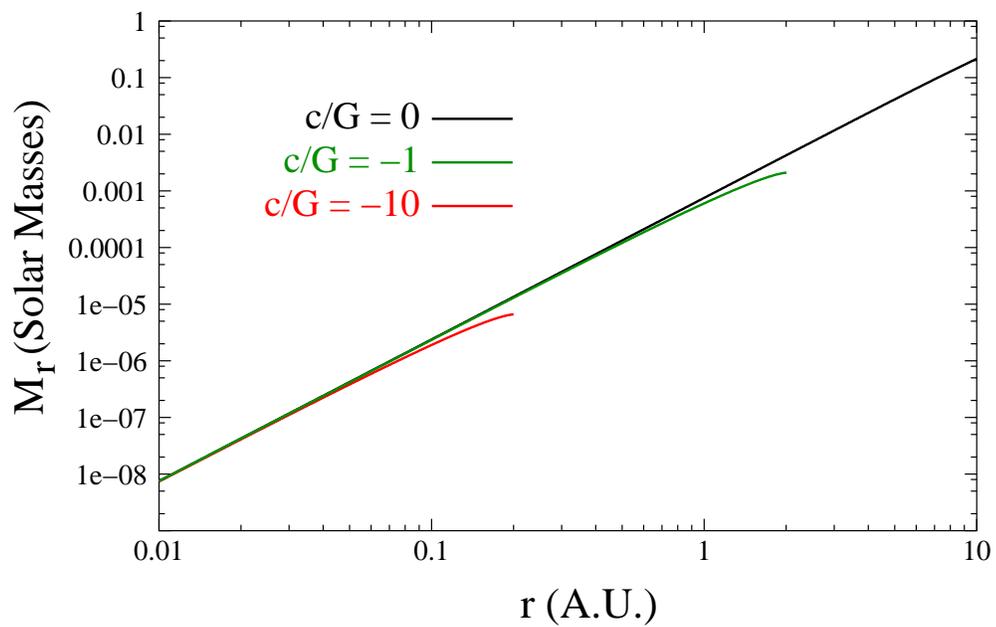}
\end{center}
\caption{The mass contribution of pseudoscalars, $M_r$, as a function of the
distance from the center of the sun. 
}
\label{fig:density}
\end{figure}

In fig. \ref{fig:density} 
we show results of the numerical integration of coupled equations
\ref{eq:dMr} and \ref{eq:dRhor}. All masses are expressed in units of
solar mass and distances in units of astronomical unit (AU). 
We have set the mass of the 
pseudoscalar equal to 1 meV. The results are shown for different values of
the parameter $c/G$. We integrate the equations starting from the 
core of the sun. The equations given above can be easily generalized 
for $R_\odot<r<R_{\rm core}$.
For $c/G \ge 0$ the pseudoscalars
have enough energy to escape the gravitational attraction of the
sun and hence the halo extends to infinite distance.
For values of $c/G\lesssim -1$ we find that the radius of
the pseudoscalar halo is quite small and contributes 
negligibly to the mass of the solar system. 
Hence we find that for a wide range of parameters
we can evade all the astrophysical bounds on light pseudoscalars. 

For $c/G\ge 0$, we find that $M_r>> M_\odot$ for $r>> 1 $ AU. 
Very large values of $M_r$ are clearly unphysical 
and suggest that in this case the pseudoscalars never reach equilibrium. In
such cases pseudoscalars may rapidly cool the star which may never reach
the main sequence. We also find that the 
large $r$ behavior, $M_r>> M_\odot$, of the density of pseudoscalars
is precisely the same 
as that required for the galactic dark matter. This raises the possibility
that the galactic dark matter might be identified with these light  
pseudoscalars. If these pseudoscalars populate the galaxy then they will 
also exert pressure on the pseudoscalars emitted by the sun. This may
also have considerable influence on the pseudoscalar density profile and will
be investigated in a future publication. 

So far we have shown that, assuming the current equilibrium conditions
inside sun's core, the pseudoscalars contribute negligibly to radiative
transport. Since all the bounds imposed in the literature consider 
perturbations around the equilibrium conditions, our analysis show that 
such bounds are not valid in the presence of self couplings of pseudoscalars. 

We next consider the approach of the star to the main sequence phase.
Since we are considering a relatively large pseudoscalar photon coupling
the pseudoscalar flux can be significant even when the star has not 
reached the main sequence. When the temperature is lower
than the current temperature the pseudoscalar flux and hence their density
inside the core will be smaller. Hence for a certain temperature range 
the pseudoscalars will escape freely. We consider the phase when the 
star undergoes quasi-static collapse due to its gravitational attraction.
By virial theorem the total energy of a
star in equilibrium is equal to half its potential energy. Hence 
its energy loss is limited by virial theorem and pseudoscalar
production cannot change its evolution drastically.
We point out that the star will evolve quasi-statically even in the presence
of pseudoscalars since the processes
leading to equilibrium proceed at a much faster rate in comparison to the
production rate of pseudoscalars.
As the core temperature  
increases, the rate of photon and hence pseudoscalar production also increases.
Once the photon production rate reaches values close to what is observed
today, pseudoscalars will start accumulating and their contribution to
radiative transport will rapidly decrease.

\section{Conclusions} 
In conclusion we find that it is possible to evade all astrophysical
limits on pseudoscalar photon coupling if the self coupling 
of pseudoscalars is sufficiently strong. The pseudoscalars start accumulating
inside the sun due to scattering with other pseudoscalars. The energy per
pseudoscalar degrades due to higher order processes as well as by 
energy loss
due to scattering on electrons, nucleons and pseudoscalars. The pseudoscalar
number density eventually reaches steady state due to conversion
of pseudoscalars back into photons. We find that the mean free path of
these particles inside the core is much smaller in comparison to the mean
free path of photons. Hence they contribute almost negligibly to radiative
transport. In our analysis we have used the pseudoscalar coupling and
mass extracted by the PVLAS experiment. Our results show that the results
of the PVLAS experiment are consistent with astrophysical limits if we 
allow strong self couplings. 

\bigskip
\noindent
{\bf Acknowledgements:} 
We thank 
Pasquale D. Serpico and Javier Redondo for a useful communication.

\end{document}